\documentclass{sigplanconf}

\usepackage{amsmath,amssymb, latexsym}
\usepackage[noend]{algorithmic}
\usepackage{algorithm}
\usepackage{pstricks}
\usepackage{graphicx}
\usepackage{xspace} 
\usepackage{subfigure}
\usepackage{epic}
\usepackage{eepic}
\usepackage{alltt}
\usepackage{listings}
\usepackage{multicol}
\usepackage{fancyvrb}
\usepackage{float}

\def\url{}
\def\@envspa{\hspace{0.3em}}
\def\@sa{\hspace{-0.2em}}
\def\@sb{\hspace{0.5em}}
\def\@sc{\hspace{-0.1em}}
\def\ba{{\sf BugAssist}\xspace}
\newcommand{\subsc}[1]{_{#1}}

{\itshape}{}
{\bfseries\upshape}{\itshape}
\newtheorem{@protheo}{Theorem}

{\bfseries\upshape}{\rm}
{\bfseries\upshape}{\itshape}
\floatstyle{ruled}
\newfloat{code}{thp}{lop}
\floatname{code}{Program}
\newcounter{ex}
\stepcounter{ex}
\newbox\subfigbox


\makeatletter

\begingroup \catcode `|=0 \catcode `[= 1
\catcode`]=2 \catcode `\{=12 \catcode `\}=12
\catcode`\\=12 |gdef|@xcomment#1\end{comment}[|end[comment]]
|endgroup

\def\@comment{\let\do\@makeother \dospecials\catcode`\^^M=10\def\par{}}

\def\begincomment{\@comment\@xcomment}

\makeatother

\newenvironment{comment}{\begincomment}{}

\def\true{{\tt true}}
\def\false{{\tt false}}

\newcommand{\ASSUME}{{\mathsf{assume}}}
\newcommand{\program}{{\mathcal{P}}}
\newcommand{\Trans}{{\mathcal{T}}}
\newcommand{\val}{v}
\newcommand{\locs}{{\mathcal{L}}}
\newcommand{\TF}{{\mathsf{TF}}}

\def\set#1{{\{ #1 \}}}
\def\tuple#1{{\langle #1 \rangle}}

\newcommand{\dbrkts}[1]{[\![#1]\!]}
\newcommand{\refine}[1]{\Delta}
\newcommand{\bug}[1]{\underline{\textcolor{red}{\textbf{\emph{#1}}}}}
\newcommand{\codecomment}[1]{\emph{#1}}

\begin{document}
\title{ Cause Clue Clauses:\\
Error Localization using Maximum Satisfiability }

\authorinfo{ Manu Jose and Rupak Majumdar}{ University of California, Los Angeles\\
MPI-SWS, Kaiserslautern}
{\tt \{manu,rupak\}@mpi-sws.org}


\maketitle

\begin{abstract}
Much effort is spent everyday by programmers in trying to reduce long, failing execution
traces to the {\em cause} of the error.
We present a new algorithm for error cause localization based on a reduction to the maximal
satisfiability problem (MAX-SAT), which asks what is the maximum number of clauses of a Boolean
formula that can be simultaneously satisfied by an assignment.
At an intuitive level, our algorithm takes as input a program and a failing test, and
comprises the following three steps.
First, using symbolic execution, we encode a trace of a program as a Boolean {\em trace formula}
which is satisfiable iff the trace is feasible.
Second, for a failing program execution (e.g., one that violates an assertion or a post-condition),
we construct an {\em unsatisfiable} formula by taking the trace formula and additionally asserting
that the input is the failing test and that the assertion condition does hold at the end.
Third, using MAX-SAT, we find a maximal set of clauses in this formula that can be satisfied
together, and output the complement set as a potential cause of the error.

We have implemented our algorithm in a tool called \ba  
for C programs. 
We demonstrate the surprising effectiveness of \ba on a set of benchmark examples 
with injected faults,
and show that in most cases, \ba can quickly and precisely isolate the exact few lines of code
whose change eliminates the error.
We also demonstrate how our algorithm can be modified to automatically suggest fixes for common
classes of errors such as off-by-one.

\end{abstract}

\section{Introduction}

A large part of the development cycle is spent in debugging, where the programmer
looks at a long, failing, trace and tries to localize the problem to a few lines
of source code that elucidate the cause of the problem.
We describe a novel algorithm for {\em fault localization} for software.
The input to our algorithm is a program, a correctness specification (either a post-condition,
an assertion, or a ``golden output''), and a program input and corresponding execution 
(called the {\em failing execution})
that demonstrates the violation of the specification.
The output is a minimal set of program statements such that there exists a way to replace these
statements such that the failing execution is infeasible.

Internally, our algorithm uses symbolic analysis of software based on Boolean satisfiability,
and reduces the problem to {\em maximum Boolean satisfiability}.
It takes as input a program and a failing test case and 
performs the following three steps.
First, 
it constructs a symbolic {\em trace formula} for the program path executed by the test input.
This is a Boolean formula in conjunctive normal form such that the formula is satisfiable iff the program
execution is feasible (and every satisfiable assignment to the formula correspond to the sequence
of states in a program execution).
The trace formula construction proceeds identically to symbolic execution or bounded model checking
algorithms \cite{King76,Biere99,cbmc,GKS05}.

Second, it extends the trace formula by conjoining it with constraints that ensure the initial state satisfies  
the values of the failing test and the final states satisfy the program post-condition that was
failed by the test.
The extended trace formula essentially states that starting from the test input and executing the program
trace leads to a state satisfying the post-condition. 
Obviously, the extended trace formula for a failing execution must be unsatisfiable.

Third, it feeds the extended trace formula to a {\em maximum satisfiability solver}.
Maximum satisfiability (MAX-SAT) is the problem of determining the maximum number of clauses of a given
Boolean formula that can be satisfied by any given assignment.
Our tool computes a maximal
set of clauses of the extended
trace formula that can be satisfied, and take the complement of this set as a candidate set of clauses
that can be changed to make the entire formula satisfiable.
Since each clause in the extended trace formula can be mapped back to a statement in the code,
this identifies a candidate localization of the error in terms of program statements. 
Note that there may be several minimal sets of clauses that can be found in this way, and we enumerate
each minimal set as candidate localizations for the user.
In our experiments, we have found that the number of minimal sets enumerated in this way remains small.

More precisely, our algorithm uses a solver for {\em partial} MAX-SAT. 
In partial MAX-SAT, the input clauses can be marked {\em hard} or {\em soft}, and the MAX-SAT instance
finds the maximum number of soft clauses that can be satisfied by an assignment 
which satisfies every hard clause.
In our algorithm, we mark the input constraints (that ensure that the input is a failing test) as well
as the post-condition are hard.
This is necessary: otherwise, the MAX-SAT algorithm can trivially return that changing an input or
changing the post-condition can eliminate the failing execution.
In addition, in our implementation, we group clauses arising out of the same program statement together,
and keep the resulting MAX-SAT instance small.

We have implemented our algorithm in a tool called \ba for fault localization of C programs.\footnote{
The tool, Eclipse plugin, and testcases can be downloaded from our web page 
\url{http://bugassist.mpi-sws.org}.
}
\ba takes as input a C program with an assertion, and a set of failing test cases,
and returns a set of program instructions whose replacement can remove the failures.
It builds on the CBMC bounded model checker for construction of the trace formula and
an off-the-shelf MAX-SAT solver \cite{msuncore} to compute the maximal set of satisfied clauses.
We demonstrate the effectiveness of \ba on 5 programs from  Siemens set of benchmarks with injected faults
\cite{testsuite}. The TCAS program in the testsuite is run with all the faulty versions in detail to 
illustrate the completness of the tool.
In each case, we show that \ba can efficiently and precisely determine the exact (to the human)
lines of code that form the ``bug''. The other 4 programs are used to show the scalability of the tool 
by using error trace reduction methods for real world programs. 

We can extend our algorithm to suggest {\em fixes} for bugs automatically, by noticing that 
the MAX-SAT instance can be used not only to localize problems, but also to suggest alternate
inputs that will eliminate the current failure.
In general, this is an instance of Boolean program synthesis, and the cost of the search can
be prohibitive.
However, we have experimentally validated that automatic suggestions for fixes is efficient 
when we additionally restrict the search to common classes of programmer errors, such as replacement
of comparison operators (e.g., $<$ by $\leq$) or off-by-one arithmetic errors.
For these classes of systems, \ba can automatically create suggestions for program changes that
eliminate the current failure.

Error localization is an important step in debugging, and improved automation for error localization
can significantly speed-up manual debugging and significantly improve the usability of automatic
error-detection tools (such as model checkers and concolic testers).
Based on our implementation and experimental results, we feel \ba is a simple yet precise technique
for error localization that effectively leverages efficient SAT solving techniques for error detection
and applies them to error localization.

\smallskip
\noindent
{\bf Related Work.}
Fault localization for counterexample traces has been an active area of research
in recent years \cite{ BallNaikRajamani03,Groce04,Groce06,Griesmayer,renieres,QiDarwin09}.
Most papers perform localization based on multiple program runs, both successful and failing,
and defining a heuristic metric on program traces to identify locations which separate
failing runs from successful ones.

Griesmayer et al.\cite{Griesmayer} gives a fault localization algorithm
for C programs by constructing a modified system that allows a given number of expressions
to be changed arbitrarily and using the counter example trace from a Model
Checker. This requires instrumenting each expression $e_i$ in the program with
$(\mathtt{diag} == i ? \mathtt{nondet}():e_i)$, where $\mathtt{diag}$ is a non deterministic variable
and $\mathtt{nondet}()$ is a new variable with the size equal to that of $e_i$. 
The number of diagnosis variables is equal to the number of components 
that are faulty in the program and need to be analyzed before creating the 
modified system.  So each expression in the program requires a
new variable in the modified system along with the diagnosis variables which 
could blow up the size of the instrumented program under consideration. 
In this work  we  avoid these drawbacks using 
selector variables and efficient {MAX-SAT} instance formulation 
using clause grouping technique. 

Many existing work \cite{renieres,Groce06,Zeller02} on fault localization  uses 
the difference between faulty trace and a number of successful traces. 
The  work of Ball at el. \cite{BallNaikRajamani03} use multiple calls 
to a model checker and compare the counterexamples to a successful trace. 
The faults are those transitions that does not appear in a correct trace. 
Our approach does not require comparing the traces or a successful 
run of the program as benchmark.  We report the exact locations where 
the bug could be corrected instead of a minimal code fragment or a fault
neighbor location.  

An alternate approach to reduce the cognitive load of debugging is {\em delta debugging}
\cite{Zeller02}, where multiple runs of the program are used to minimize the ``relevant''
portion of the input.
We believe our technique is orthogonal to delta-debugging and its variants, and can
be composed profitably.

While we describe our algorithm in pure symbolic execution terms, our algorithm fits in very well
with concolic execution \cite{GKS05,SMA05,CadarDE08}, where symbolic constraints are generated while
the concrete test case is run. 
Our motivation for using CBMC was the easy integration with MAX-SAT solvers, but in our implementations,
we performed some optimizations (such as using concrete values for external library calls in the trace formula
and constant-folding input-independent parts of the constraints) similar to concolic execution.

The motivation to use unsatisfiability cores is their recent success in hardware
circuit design debugging described in Safarpour et al. 
\cite{Safarpour07,safarpour09}.  {MAX-SAT} based debugging  is used as a 
framework for debugging  gate level {VLSI} circuits. 
Unsatisfiability cores have also been  used to pin point  over-constrains in 
declarative models \cite{DeclarativeMaxSAT03}.

\section{Motivating Example}

\begin{code}  
\begin{Verbatim}[commandchars=\\\{\},
               codes={\catcode`$=3\catcode`^=7}]  
int Array[3];
int testme(int index)
  \{
  .   ..................
  1  \bug{if ( index != 1)} \codecomment{/* Potential Bug 2 */}
  2     index = 2;
  3  else
  4     \bug{index = index + 2;} \codecomment{/* Potential Bug 1 */}
  .   ...................
  .   ...................
  .
  5  i = index;
  6  return Array[i]; //assert(i >= 0 && i < 3)
  \}
\end{Verbatim}
\caption{A simple example.} 
\label{fig.eg}
\end{code}

We start with an informal description of \ba. 
Consider the function {\tt testme} in Program \ref{fig.eg} 
which returns  a value at a new location from an array of size 3. 
The function takes in two arguments: the array itself and the 
current index value. 
The function does some computation on  the current index value (shown in lines~1--4) 
to find the new index and returns the value at new index in line~6.
The array dereference on line~5 generates implicit assertions
about the array bounds shown in line~6.

The program has a bug.
If the input $\mathtt{index}$ is equal to $1$, then the else-branch sets
$\mathtt{index}$ to $3$, and the subsequent array dereference on line~6
is out of bounds.
Testing the program with this input 
will find the bug, and return a program trace that shows the array bounds violation at the end.
%
But testing or model checking returns a {\em full execution} path, including details irrelevant to the
specific bug, and do not give the 
reason for failure, or the cause of the bug.
The localization algorithm in \ba helps to nail down the 
issue to a few potential bug locations in the program where 
the correction has to be made. 

\ba works as follows.
Starting with the test input $\mathtt{index} = 1$ and the corresponding program trace:
\[
\ASSUME(\mathtt{index} = 1); \mathtt{index} = \mathtt{index} + 2; \mathtt{i} = \mathtt{index};
\]
it first constructs a symbolic {\em trace formula} $\TF$ encoding the execution trace:
\[
\TF \equiv \mathtt{index}_1 = 1 \wedge \mathtt{index}_2 = \mathtt{index}_1 + 2  \wedge \mathtt{i} =\mathtt{index}_2
\]
We assume that integers and integer operations are encoded in a bit-precise way, 
and without loss of generality,
the trace formula is a Boolean formula in conjunctive normal form.
We omit the standard encoding from imperative programs to Boolean formulas (see, e.g., \cite{cbmc}). 

Clearly, at the end of the trace, the assertion
\[
\mathtt{i} < 3  
\]
does not hold.
Consider now the formula
\[
\Phi\equiv\underbrace{\mathtt{index}_1 = 1}_{\mathrm{test\ input}} \wedge \underbrace{\TF}_{\mathrm{trace\ formula}} \wedge \underbrace{\mathtt{i} < 3}_{\mathrm{assertion}}
\]
which is unsatisfiable.
Intuitively, the formula captures the run of the program starting with the error-inducing test input, and
asserts that the assertion holds at the end (a contradiction, by choice of the input).

We convert ${\Phi}$ to conjunctive normal form (CNF) and feed it to 
a partial MAX-SAT solver \cite{msuncore}. 
A partial MAX-SAT solver takes as input a Boolean formula in CNF,
where each clause is marked ``hard'' or ``soft'' and returns the maximum
number of soft clauses (as well as a subset of clauses of maximum cardinality)
that can be simultaneously satisfied by an assignment
satisfying all the hard clauses.
In case of $\Phi$, we make the constraints coming from the test input ($\mathtt{index} = 1$)
and the assertion ($\mathtt{i} < 3$) as hard, and leave the clauses in the trace formula soft.
Intuitively, we ask, given that the input and the assertion are fixed, which parts of the trace
formula are consistent with the input and the assertion?
The partial MAX-SAT solver then tries to 
find a set of soft clauses of maximum cardinality which can be simultaneously satisfied while
satisfying all the hard clauses. 
The Complement of a set of maximum satisfiability clauses (CoMSS) gives a set of soft
clauses of minimum cardinality whose removal would make ${\Phi}$  satisfiable, i.e., consistent
with the view that the test input does not break the assertion.
We use this set as potential locations of the program error. 

In addition, by grouping together clauses arising out of the same program statement,
we can map the clauses back to the lines of the program. 
Using clause grouping, described in Section~\ref{sec-prelim}, each line in the program 
is mapped to a bunch of its soft clauses which are enabled and disabled 
simultaneously.

In our example, the hard and soft clauses are:
\begin{align*}
\mathrm{Hard}: & \mathtt{index} = 1 \wedge \mathtt{i} < 3\\
\mathrm{Soft}: & \TF
\end{align*} 
MAX-SAT returns that a possible CoMSS maps to the line~4 in the program.
This is the unsatisfiable core whose removal or correction can 
satisfy the formula  ${\Phi}$. 
We claim that is a potential error location for the program and a fix would be
to change the constant to any integer less than 2 and greater than -2. 

Suppose this is not where  programmer wants to make a correction
and require other locations where he could fix the bug.
We iterate by making another call to MAX-SAT, but this time make clauses arising out of line~4
hard, i.e., asking the MAX-SAT for possible CoMSS where line~4 is kept unchanged.
This reveals another potential bug location in the code.
 We repeat this process until MAX-SAT 
gives the problem to be unsatisfiable and no more clauses can be removed to make this 
problem satisfiable. The error locations reported by \ba are underlined  in
 Program  \ref{fig.eg}. On a closer look, these are all the places where the correction can be made.
 Either changing the constant value at line~4  or the conditional
 statement at line~1 can  fix the program. 
\ba is available as an Eclipse plug-in, making it easy for the programmer to interactively find potential
error points.

Notice that our technique is stronger than simply taking the backward slice of the program trace,
and gives fine-grained information about potential error locations.
The backward slice for this trace contains all the lines 1,4, and 5.
Our algorithm returns lines 1 and 4 separately as potential error locations.

So far we have focused on error localization.
The methodology can be modified to suggest program repairs as well. 
Intuitively, the fault localization returns a set of program commands that
are likely to be wrong.
One can then ask, what are potential replacements to these commands that fixes the
error?
In general, the space of potential replacements is large, and searching this space
efficiently is a difficult problem of program synthesis \cite{Bodik05,vs3}.
Instead, we take a pragmatic approach and look for possible fixes for common programmer errors.

Specifically, we demonstrate our idea by fixing 
``\emph{off by one}"  errors. In this example, the error
occurs due to accessing an out of bound array element by one.
When \ba comes back with  line~4 as a potential bug location, 
we try to ``fix'' the bug by changing the constant whose new value is one off its 
current value. 
So we change the value 2 in this line to 3 or 1 and check if either of these 
values satisfy the properties.
This involves modifying the trace formula appropriately and checking if the
failing program execution becomes infeasible with either change. 
So in this case we create two programs with new constants 
 at line~4 as follows. 
\begin{align*}
   Program 1:   \ \ index = index + 3 \ \   \times \\
   Program 2:   \ \ index = index + 1 \ \   \surd 
\end{align*} 
The new value 1 ensures that the error path is infeasible,
and this can be used as a suggestion for repair for the program. 
The same procedure can be used to check for 
operator errors like use of plus instead of 
minus, division instead of multiplication,
performing assignment instead of
equality test, etc., which are common programmer error patterns.
\label{example.sec}

\section{Preliminaries}
\label{sec-prelim}

\subsection{Programs: Syntax and Semantics}

We describe our algorithm on a simple imperative language based on control-flow graphs.
For simplicity of description, we omit features such as function calls or pointers. These are handled
by our implementation.

A {\em program} $G = (X,\locs, \ell_0, \Trans)$ consists of a set $X$
of Boolean-valued variables, a set $\locs$ of {\em control locations},
an initial location $\ell_0\in\locs$ and a set $\Trans$ of {\em
  transitions}.
Each transition $\tau \in \Trans$ is a tuple $(\ell,\rho,\ell')$ where
$\ell$ and $\ell'$ are control locations and $\rho$ is a constraint
over free variables from $X\cup X'$, where the variables from $X'$
denote
the values of the variables from $X$ in the next state.

For a constraint $\rho$, we sometimes write $\rho(X,X')$ to denote that
the free variables in $\rho$ come from the set $X \cup X'$.

Our notation is sufficient to express common imperative programs
(without function calls): the
control flow structure of the program is captured by the graph of
control locations, and operations such as assignments $x:= e$ and
assumes $\ASSUME(p)$ captured by constraints
$x' = e \wedge \bigwedge \set{ y' = y \mid y \in X \setminus \set{x}}$ 
and $p \wedge \bigwedge \set{x' = x \mid x\in X}$
respectively.

A {\em state} of the program $\program$ is a mapping from 
variables in $X$ to Booleans.
We denote the set of all program states by $\val.X$.
A {\em computation} of the program is a sequence
$\tuple{m_0,s_0}\tuple{m_1,s_1}\ldots \in (\locs\times \val.X)^*$,
where $m_0 = \ell_0$ is the initial location, and for each
$i\in\set{0,\ldots, k-1}$, there is a transition
$(m_i,\rho_i,m_{i+1})\in\Trans$ such that $(s_i,s_{i+1})$ satisfies
the constraint $\rho_i$.

An {\em assertion} $p$ is a set of program states.
A program {\em violates} an assertion $p$ if there is some computation
$\tuple{m_0,s_0}\ldots\tuple{m_k,s_k}$ such that $s_k$ is not in $p$.
Typically, assertions can be given as language-level correctness requirements
(e.g., ``no null pointer dereference''), as programmer-specified {\sf assert}s in the code,
or as post-conditions.

\subsection{Trace Formulas}

A {\em trace} $\sigma$ is a finite sequence 
$(m_0,\rho_0,m_1), (m_1,\rho_1,m_2), \ldots, (m_{k-1},\rho_{k-1},m_{k})$ of transitions in $\Trans$ 
such that $m_0 = \ell_0$.
The trace $\sigma$ is {\em feasible} if there exists a computation 
$\tuple{m_0,s_0}\ldots\tuple{m_k,s_k}$
such that for each $i\in \set{0,\ldots,k-1}$, we have $(s_i,s_{i+1})$ satisfies $\rho_i$.

Given a trace $\sigma$, we define the {\em trace formula} $\TF(\sigma)$
as the conjunction
\begin{equation}\label{tf}
\bigwedge_{i = 0}^{k-1} \rho_i(X_i, X_{i+1})
\end{equation}
where $X_i$ is a copy of the variables in $X$ for each $i\in\set{0,\ldots,k}$
and
$\rho_i(X_i,X_{i+1})$ denotes the constraint $\rho_i(X,X')$ with the variables in $X$ substituted
by corresponding variables in $X_i$ and the variables in $X'$ substituted by corresponding variables in $X_{i+1}$.
Note that $\TF(\sigma)$ is satisfiable iff the trace $\sigma$ is feasible.

While we have described Boolean programs, a C program with finite-bitwidth data, e.g., 32-bit integers,
can be converted into an equivalent Boolean program by separately tracking each bit of the state, and
by interpreting fixed-width arithmetic and comparison operators as corresponding Boolean operations on
each individual bit.
We omit the (standard) details, see e.g., \cite{cbmc,XieAiken}.

\subsection{Partial Maximum Satisfiability}


Given a Boolean formula in conjunctive normal form, the {\em maximum satisfiability}
(MAX-SAT) problem asks what is the maximum number of clauses that can be satisfied
by any assignment \cite{maxsatbook}.
The MAX-SAT decision problem is NP-complete; note that a formula is satisfiable iff all its
clauses can be satisfied by some assignment.

The {\em partial maximum satisfiability} (pMAX-SAT) problem takes as input a Boolean formula $\Phi$
in conjunctive normal form, and a marking of each clause of $\Phi$ as {\em hard} or {\em soft},
and asks what is the maximum number of soft clauses which can be satisfied by an assignment
to the variables which satisfies all hard clauses. 
Intuitively, each hard clause must be satisfied, and we look for the maximum
number of soft clauses which may be satisfied under this constraint.

Recent years have seen a tremendous improvement in engineering efficient
solvers for MAX-SAT and pMAX-SAT. 
The widely used algorithm for MaxSAT is based on branch-and-bound search 
\cite{Li07newinference},  supported by effective lower bounding
 and dedicated inference techniques. 
Recently, unsatisfiability based MaxSAT solvers 
by iterated identification of unsatisfiable sub-formulas was proposed 
in \cite{Fu06onsolving}.
  This approach consist of identifying unsatisfiable 
sub-formulas and relaxing clauses in each unsatisfiable sub-formulas
 by associating a relaxation variable with each such clause. 
Cardinality constraints are 
used to constrain the number of relaxed clauses  \cite{msuncore,msc}. 

In addition to solving the decision problem, MAX-SAT solvers
also give a set of clauses of maximum cardinality that can be simultaneously satisfied.
The complement of these maximum satisfiable subsets (MSS) are a set of clauses 
whose removal makes the instance satisfiable(CoMSS). 
Since the maximum satisfiability subset
is maximal the complement of this set is minimal \cite{Liffiton05onfinding}.

 In this work we make use of these CoMSS which refers to the clauses whose
 removal can make the system satisfiable.
 Since we represent a C program as a boolean satisfiability problem
with constraints and properties, these coMSS are oracles for potential bug locations.

\subsection{Efficient Compilation to MAX-SAT} 


A single transition can lead to multiple clauses in the conjunctive normal
form of the trace formula.
In this section we  suggest a method to simplify the MAX-SAT problem 
by grouping together clauses arising out of a single
source-code statement.
We now give a simple way of grouping clauses arising out of the same program operation.

For each transition $\tau = (m,\rho,m')\in\Trans$, we introduce a new Boolean variable $\lambda_\tau$.
Then, we augment each clause arising out of $\rho$ with $\lambda_\tau$.
For example, suppose $(c^1_1 \vee \ldots) \wedge (c^2_1 \vee\ldots)$ is a conjunctive
normal form representation of $\rho$, then the augmented representation is
$(\lnot \lambda_\rho \vee c^1_1\vee\ldots) \wedge (\lnot \lambda_\rho \vee c^2_1\vee\ldots)$.

The augmentation with $\lambda_\rho$ has the following effect.
When $\lambda_\rho$ is assigned $\true$, the original clauses in the CNF representation
of $\rho$ must be satisfied, while when $\lambda_\rho$ is assigned $\false$,
each augmented clause is already satisfied.
This helps to enable and disable the clauses corresponding to each transition
by setting and unsetting the $\lambda_\rho$ variable respectively.
The $\lambda$-variables are called {\em selector variables}.

We use a representation of trace formulas using selector variables.
Instead of Equation~\eqref{tf} for the trace formula, we use the form:
\begin{equation}\label{aug-tf}
\underbrace{\bigwedge_{i=1}^{k-1} \mathsf{CNF}(\rho_i(X_i,X_{i+1}),\lambda_{\rho_i})}_{\TF_1} \wedge \underbrace{\bigwedge_{(\cdot,\rho,\cdot)\in\Trans}\lambda_{\rho}}_{\TF_2}
\end{equation}
where $\mathsf{CNF}(\rho,\lambda_\rho)$ denotes the augmented representation
for the CNF for $\rho$, and we label the two parts of the formula $\TF_1$ and $\TF_2$ for later
reference.
Intuitively, clauses from $\TF_1$ will be marked as hard clauses to the MAX-SAT solver,
and clauses from $\TF_2$ will be marked soft.
Thus, the MAX-SAT solver will explore the space of possible program statements whose replacement
will cause the error to go away.

Notice that we allocate a selector variable for each transition of the program, so the number of
selector variables is bounded by the size of the program. However, in a trace, the same program
transition may occur multiple times (e.g., on unrolling a loop), and there is a distinct clause for
each of these occurrences all tagged with the same selector variable.

\begin{comment} 
A program given to the Model Checker and is converted to an 
unsatisfiable boolean formula as explained in Section \ref{example.sec} to  
$\hat{\Phi} = I \wedge C \wedge P$, where $I$ is the variable assignment for 
the counter example given by model checker, C is the set of constraints for 
the program and P is the set of safety properties.
We augment the $\lambda$ variables to the clauses in $C$
as explained above and create a new partial MaxSAT instance
as below.
\begin{align*}
\hat{\Phi} & = \Phi_H \cdot \Phi_S, where \\
 & \Phi_H = I \wedge P   \bigwedge_{i=1}^{n}  c \in C_i, (\lambda_i \rightarrow c),\\
 & \Phi_S = \bigwedge_{i=1}^{n}  \lambda_i
\label{eq.pmax}   
\end{align*}
Here $n$ is the total line numbers in the program. 
It should be noted that the optimization problem for MaxSAT  is to 
find the maximum satisfiable assignment for the soft clauses in $(|\Phi_S|=n)$
and to keep the clauses in $\Phi_H$ always true. 
This basically means to enable or disable the selector variables which 
corresponds to the line numbers in the program. 
This keeps the MaxSAT optimization problem small (n)
and gets mapped to the original program lines easily. 
\end{comment}

We use the abstraction technique on transitions, which correspond to line numbers
of code in our implementation, but it is also possible
to group the clauses from modules and recursively narrow down the problem to 
a module, and then to a line.

\section{Algorithm}


We now describe the algorithm for \ba.
There are two phases of the algorithm: first, generate a failing execution (and a test
demonstrating a failing execution), and second, find a minimal set of transitions
that can render the failing execution infeasible.

\subsection{Generating Traces}

In general, any method of generating a failing execution of a program can be 
used as a starting point of our algorithm.
In our implementation, we use two approaches.
In case the program comes with a testsuite, we generate failing executions from failed tests.
In case there are no available tests, we use
{\em bounded model checking} \cite{Biere99,cbmc}
to systematically explore program executions and look for potential assertion violations.
If a failing execution is found, the bounded model checking procedure can generate a concrete
initial state that leads to the assertion violation.

\subsection{The Localization Algorithm}

\begin{algorithm}[t]
\begin{algorithmic}[1]
\ALGINPUT Program $\program$ and assertion $p$
\ALGOUTPUT Either $p$ holds for all executions or potential bug locations
\STATE $(\mathsf{test},\sigma)$ = GenerateCounterexample(P, p)
\IF{$\sigma$ is ``None''}
\STATE $\textbf{return}$ ``No counterexample to $p$ found''
\ELSE
\STATE $\Phi_H = \dbrkts{\mathsf{test}} \wedge p \wedge \TF_1(\sigma)$
\STATE $\Phi_S = \TF_2(\sigma)$
\WHILE{$\true$}
\STATE BugLoc = CoMSS($\Phi_H, \Phi_S$)
\IF{BugLoc $=$ $\emptyset$}
\STATE $\textbf{return}$ ``No more suspects''
\ELSE
\STATE  $\textbf{output}$ ``Potential bug at CoMSS BugLoc"
\STATE $\beta$ = $\bigvee \set{\lambda\subsc{i}\mid \lambda\subsc{i}\in \mathrm{BugLoc}}$
\STATE  $\Phi\subsc{S}=\Phi\subsc{S} \backslash \beta$ 
and $\Phi\subsc{H}=\Phi\subsc{H} \cup \beta$
\ENDIF
\ENDWHILE
\ENDIF
\end{algorithmic}
\caption{Localization Algorithm\label{fig.alg}}
\end{algorithm}

Algorithm \ref{fig.alg} shows the \ba localization algorithm. 
Line~1 calls the procedure to generate failing executions for the assertion.
If no failing executions are found, the procedure returns.
Otherwise, we get a concrete test case $\mathsf{test}$
as well as a program trace $\sigma$ demonstrating
the failure of the assertion.

Using the test, the failing execution, and the assertion, we construct two formulas (lines~5,6).
The formula $\Phi_H$ consists of three parts. The first part, $\dbrkts{\mathsf{test}}$,
is a formula asserting that the initial
state coincides with the test case that caused the failure.
Formally, for a program state $s$, the constraint $\dbrkts{s}$ is defined as $\bigwedge\set{ x = s(x) \mid x \in X}$.
The second part is the assertion $p$.
The third part is the first part $\TF_1(\sigma)$ of the trace formula from Equation~\eqref{aug-tf}.
The formula $\Phi_S$ is the second part $\TF_2(\sigma)$ of the trace formula from Equation~\eqref{aug-tf}.

Notice that $\Phi_H \wedge \Phi_S$ is unsatisfiable. (Intuitively, it says that if the program is run
with the test input $\mathsf{test}$, then at the end of the execution trace $\sigma$, the assertion $p$ holds.)

In subsequent calls to pMAX-SAT, clauses in $\Phi_H$ are treated as hard clauses, and clauses in $\Phi_S$ are
treated as soft clauses.
Intuitively, treating $\Phi_S$ as soft clauses enables us to explore the effect of changing each subset of transitions
to see if the failing transition can be made infeasible.

The search for localizations is performed in the $\textbf{while}$ loop of lines~7--14.
During each iteration of the while loop, we call the 
pMAX-SAT solver and get a CoMSS for the current $(\Phi_H, \Phi_S)$ pair. 
Each of these clauses returned by  CoMSS gives potential bug locations in the code,
and is output to the programmer.

Whenever we report a potential bug, we add a hard blocking clause for the corresponding CoMSS, 
so that in subsequent iterations, this CoMSS is not explored again as a potential cause of error.
In many of our experiments, the CoMSS returns a single $\lambda_\rho$ clause as the 
indicator of error. In general, it  returns more than one selector variable which 
indicates that  the program cannot be fixed by changing any one line but must be
changed at multiple locations. (This does happen in experiments.)
Adding each of these $\lambda_\rho$ variables as a new hard clause blocks the occurrence 
of these clauses in a different clause  combination.  
To avoid this problem, we compute a blocking clause $\beta$ (lines~13)
and make the blocking clause hard.
For example, suppose the coMSS returned 
$BugLoc={\lambda_1, \lambda_2, \ldots, \lambda_k}$. This means that the bug
can be fixed by making simultaneous changes to these $k$ locations.
In the next iteration, we add a new hard clause
$(\lambda_1\vee \ldots \vee \lambda_k)$ which ensures that
this particular CoMSS is not encountered again, but other combinations of these
locations are still allowed.
 
\subsection{Dealing with Multiple Locations}

The  \ba returns multiple locations where a correction is possible.
The experimental results in section \ref{exp.1} shows that the number of 
potential error locations returned is quite small and in most cases 
the exact bug location is reported using a single failing execution. 
However, for reliabilty and further refinement of bug locations, 
we use a ranking machanism for bug locations by 
running the \ba algorithm repeatedly with  different failing program traces and ranking
the bug locations based on their frequency of appearance in each of these runs.
While using a  model checker for  counter example traces, it gets the  variable 
assignment for a SAT formula it created for the program.
By changing the order of variables in the SAT algorithm \cite{dpll} or by doing 
a random restart of the solver we can efficiently get a new counter example 
trace for the same variables. Running \ba  with these new value 
gives a another bunch of potential bug locations. Repeating this process and 
ranking the bug locations narrow down the search to a few lines in the program.

\section{Extensions}

We now describe two extensions to the basic algorithm.

\subsection{Extension 1: Automated Repair}

\ba can be used for automated repair of programs, as it
boils down the problem to a few potential bug locations  in the code.
After analysing the problem lines, we get an idea of the kind of error that
could have happened. For example, if there is a constant in the line we could
try to synthesize a new constant which can fix the code \cite{Griesmayer} or 
if there is an operator, changing the operator might be a repair for the bug.
We demonstrate this capability by fixing  "Off-By-One" \cite{wiki:offbyone}
 errors in the program. They are a common logical error involving discrete
equivalent of a boundary condition. Usually programmer forgets that
a sequence starts at zero rather than one (e.g. array indices in many 
languages like {C}, {C++}). It is also caused during boundary check 
conditions by using a $<$ instead of $\leq$ or viceversa.


During the code parsing phase, we mark the lines which has constants in them.
After running \ba on the code, it gives the potential correction locations and
if that is a marked line we assign  $^{\tiny +/-}1$ value to the constant
 in the code and ask if the new values can satisfy the properties.

The repair procedure is given in Algorithm \ref{fig.offbyone}.
In line~1 the LocalizationProcedure is called to get the potential bug locations.
The function $GetConst(i)$ checks if  line~i has a constant in it, if so it 
returns the value  to $\kappa$. We change the constant
$\kappa$ at line~i and creates two new programs $\program\subsc{1}'$ and 
$\program\subsc{2}'$ each with one off $\kappa$. 
The lines~6 and 8 check if the new programs contain an error trace.
If one of those return an empty counter example it is  declared as a repair to the 
buggy version of $\program$.

\begin{algorithm}[h]
\begin{algorithmic}[1]
\ALGINPUT Buggy Program $\program$ and assertion $p$
\ALGOUTPUT Either Fixed program  or no Off-By-One error.
\STATE BugLoc = LocalizationProcedure($\program$,$p$)
\FOR{$\textbf{all}$  $\lambda\subsc{i}\in$ BugLoc}
\IF{($\kappa=GetConst(i)$) $\neq$ $\emptyset$}
\STATE $\program\subsc{1}' = (\program \backslash \kappa) \cup (\kappa+1)$ 
\STATE $\program\subsc{2}' = (\program \backslash \kappa) \cup (\kappa-1)$ 
\IF{GenerateCounterExample($\program\subsc{1}'$,$p$) $=$ $\emptyset$} 
\STATE $\textbf{return}$ $\program\subsc{1}'$
\ENDIF
\IF{GenerateCounterExample($\program\subsc{2}'$,$p$) $=$ $\emptyset$}
\STATE $\textbf{return}$ $\program\subsc{2}'$
\ENDIF
\ENDIF
\ENDFOR
\STATE  $\textbf{output}$ ``Off-By-One error not found"
\end{algorithmic}
\caption{The Off-By-One Repair Algorithm}
\label{fig.offbyone}
\end{algorithm}

\subsection{Extension 2: Debugging Loops}

The bugs in loop body can be burdensome to fix as they might be hidden
in initial  iterations  and visible afterwards. 
The usual model checker methodology to verify properties  is by
 loop unwinding which duplicates the loop body $\eta$ times, where $\eta$ is 
the unwinding limit. The programmer would
be interested in knowing the iteration at which the assertion is violated
to get a better idea about the cause of the error. We suggest a method
to catch the potential iteration of the loop where the bug appeared first.

This is achieved using clause grouping and assigning weights to the soft clauses
in the {pMAX-SAT} instance. Each time  a loop body is duplicated
(till bound $\eta$) we create a new selector variable.
For example, for a transition $\tau = (m,\rho,m')\in\Trans$ in the
while loop body, during $\kappa^{th}$ unwinding we augment each clause
arising out of  $\rho$ with $\lambda_{\tau}^{\kappa}$. We add
these selector variables as soft clauses to the {MAX-SAT} instance as before,
 but assign a weight as follows
\begin{equation}
\forall_{\kappa=1}^{n}Weight(\lambda_{\tau}^{\kappa}) = \alpha+\eta-\kappa
\end{equation}
where $\alpha$ is the default weight for soft clauses. This make sure that
the clauses corresponding to the initial iterations of the loop gets 
a higher weightage.  
The weights assigned to the soft clauses in the {pMAX-SAT} is the 
penalty that has to be paid to falsify the clauses. The solver extracts the 
{CoMSS} in such a way that the least iteration clauses are picked first
as  they weigh more than the latter iterations variables.
This helps to  pin-point the initial iteration  
of the loop which can reproduce  the failure.

\section{Experimental Results}

\begin{figure}[htb]
\includegraphics[width=0.5\textwidth]{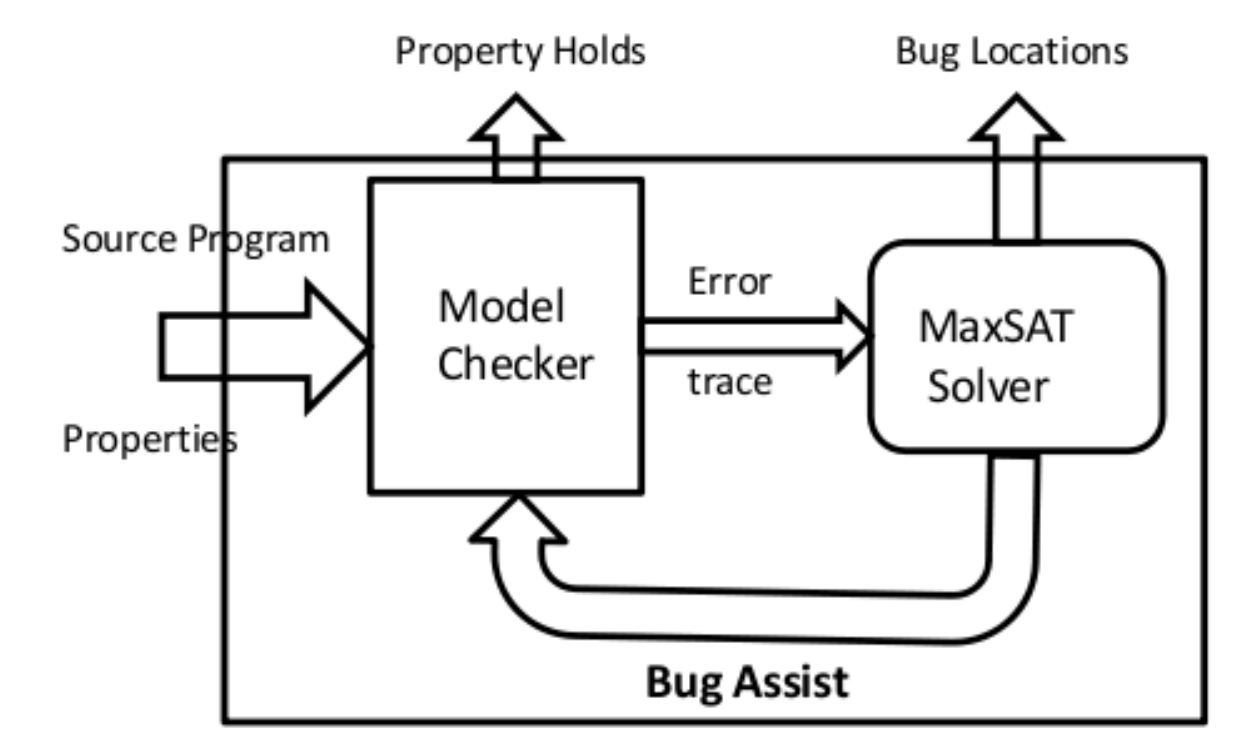}
\caption{Basic Flow Diagram.}
\label{fig.flow}
\end{figure}
\setlength{\tabcolsep}{2pt}
\begin{table*}[htb]
\centering
\caption{Results of running \ba on TCAS task of the Siemens Test Suite.}
\label{tcas.tab}
\begin{tabular}{||c|c|c|c|c|c|c||c||c|c|c|c|c|c|c|| }
\hline
\hline
Version&TC\#&Error\#&Detect\#&Size& Run&Error &--& Version&TC\#&Error\#&Detect\#&Size\%&Run&Error \\
&&&&Reduc\%& Time&Type && &&&&Reduc\%&Time&Type \\
\hline
v1 & 132 & 1 & 132 & 8.6 &0.016 & op  & & v21 & 16& 1 &  16 & 8.6 &0.108  &op\\
v2 &  69 & 1 &  69 & 4.6 &0.068 & const  & &v22 & 11& 1 &  11 & 5.7&0.056 & code   \\
v3 &  23 & 1 &  13 & 9.8 &0.096 & op &  &v23 & 42& 1 &  41 &6.3 &0.100  & code \\
v4 &  26 & 1 &  26 & 9.2 &0.104 & op &  &v24 & 7 & 1 &   7 &8.6&0.092  & op\\
v5 &  10 & 1 &  10 & 8.6 & 0.120& assign&  &v25 & 3 & 1 &   3 & 6.9& 0.068  &code \\
v6 &  12 & 1 &  12 & 8.6 &0.108 & op&   &v26 & 11& 1 &  11 & 9.2& 0.108 & addcode \\
v7 &  36 & 1 &  36 & 9.2 &0.072 & const&   &v27 & 10& 1 &  10 & 10.9& 0.108 & addcode \\
v8 &  1  & 1 &  1  & 8.6 &0.112 & const&   &v28 & 76& 1 &  58 & 5.7& 0.080 & Branch \\
v9 &  9  & 1 &  9  & 5.2 &0.092 & op&   &v29 & 18& 1 &  14 &5.7 &0.092&  code\\
v10 & 14& 2 &  14 & 9.2 &0.136  &op&  &v30 & 58& 1 &   58& 5.7& 0.064 & code \\
v11 & 14& 2 &  14 & 6.3 &0.080 & op &  &v31 & 14& 2 &  14 &10.9 & 0.008  & addcode\\
v12 & 70& 1 &  48 & 9.2 &0.164 &  op&  &v32 & 2 & 2 &   2 & 10.9& 0.004 & addcode\\
v13 &  4& 1 &  4  & 9.2 & 0.080 & const& &v34 & 77& 1 &  77 &8.6 &0.100 & op \\  
v14 & 50& 1 &  50 & 8.1 & 0.028 & const&  &v35 & 76& 1 &  58 &5.7 &0.060 &  code\\
v15 & 10& 3 &  10 & 7.5 & 0.104 & const&  &v36 &126& 1 &  126 &2.9 & 0.024 & op \\
v16 & 70& 1 &  70 & 9.2 & 0.104 & init &  &v37 & 93& 1 &  93 &8.6 &0.040  & index\\ 
v17 & 35& 1 &  35 & 9.2 & 0.096 &init  &  &v39 & 3& 1 &  3 &6.9 &0.088 &  op\\ 
v18 & 29& 1 &  29 & 6.9 &0.124  & init &  &v40 & 126& 2 & 126 & 6.3 &0.088 &  assign\\
v19 & 19& 1 &  19 & 9.2 & 0.112 & init & &v41 & 20& 1 &  20 & 8.6& 0.120&  assign\\
v20 & 18& 1 &  18 & 9.2 &0.120  &op&     & -- & --&-- &--   &--  &--    &  \\ 
\hline
\hline
\end{tabular}
\end{table*}

\begin{figure}[htb]
\label{tcas.v2}
{\scriptsize
\begin{Verbatim}[commandchars=\\\{\} ]
1 int Inhibit_Climb () \{
2   \bug{return (Climb_Inhibit?Up_Sep+300:Up_Sep);}
3   \codecomment{/*return (Climb_Inhibit?Up_Sep+100:Up_Sep);*/}
4 \}
5 int Non_Crossing_Climb() \{
6    upward_preferred=Inhibit_Climb()>Down_Sep;
7    if (upward_preferred) \{
8         result = !(Own_Below_Threat()) || 
                   (!(Down_Sep >= ALIM()));  \}
9      else\{
10       result =  (Cur_Vertical_Sep >= 100)
                   && (Up_Sep >= ALIM());  \}
11     return result;
12 \}
13 int Non_Crossing_Descend() \{
14    \bug{upward_preferred=Inhibit_Climb()>Down_Sep;}
15    \bug{if (upward_preferred)} \{
16      \bug{ result = Own_Below_Threat() &&}
              \bug{(Cur_Vertical_Sep >= 100) &&} 
                     \bug{(Down_Sep >= ALIM());}\}
17     else\{
18       result = !(Own_Above_Threat()) ||
               ((Own_Above_Threat()) && 
               (Up_Sep >= ALIM())); \}
19     \bug{return result;}
20 \}
21  int alt_sep_test() \{   
22      enabled = true; \codecomment{/*conditions omitted*/}
23      alt_sep = UNRESOLVED;
24      if (enabled) \{
25         need_upward_RA=Non_Crossing_Climb()&& 
                             Own_Below_Threat();
26         \bug{need_downward_RA=Non_Crossing_Descend()} 
                             \bug{&& Own_Above_Threat();}
27         \bug{if (need_upward_RA && need_downward_RA)}
28             alt_sep = UNRESOLVED;
29         else if (need_upward_RA)
30             alt_sep = UPWARD_RA;
31         else if (need_downward_RA)
32             alt_sep = DOWNWARD_RA;
33      \}
34      \bug{return alt_sep;}
35 \}
36 int main() \{\codecomment{/*inputs omitted*/}
37    assert(alt_sep_test() == DOWNWARD_RA);
38 \}
\end{Verbatim}
\caption{A Sample TCAS code with declarations and several code fragments 
omitted. All bug locations identified are underlined, original code at line \codecomment{3}; 
mutation at line \codecomment{2}.}}
\end{figure}

We demonstrate the capability of the tool in this section by 
showing the results from running few programs from the
Siemens test suite \cite{testsuite}. The Siemens test suite is widely
used in the literature for bug localization study \cite{Griesmayer, renieres}.
In section \ref{exp.1}  we analyse a simple program  TCAS task \cite{tcas} in depth
 and in section \ref{exp.2} we illustrate the scalability of our method using more
complex examples.

Figure \ref{fig.flow} gives an overview of the implementation of \ba.
We used CBMC \cite{cbmc} as the model checker for generating failing traces
and test inputs.
Tests can also be fed directly. 
CBMC is a Bounded Model Checker for ANSI-C and C++ programs.  
For solving the {pMAX-SAT} instances, we 
used the \emph{Maximum Satisfiability with UNsatisfiable COREs} (MSUnCORE) tool  
\cite{msuncore}, which can handle large and complex weighted partial MaxSAT problems.
The off-by-one error fix was synthesized using 
the MiniSAT2 \cite{MiniSAT2} SAT engine.
All  our experiments are preformed on 
an 3.16 GHz Intel Core 2 Duo CPU with 7.6 GB RAM.

\subsection{TCAS Experiments}
\label{exp.1}
The TCAS task of the Siemens test suite constitutes an
aircraft collision avoidance system. It consists of 173 lines
of code. The authors have created 41 versions of the program by 
injecting one or more faults. Their goal was to introduce faults that were 
as realistic as possible, based on their experience with real programs. 
We refer to the versions as ``v1" to  ``v41". 
The suite also contains 1600 test cases which are valid inputs for the program.

We created the golden outputs
for these 1600 test cases by running the original version of 
the program. Then for each of the faulty versions, we ran
those 1600 test vectors and matched with the golden outputs to 
segregate the failing test cases. Since the program does not
 contain a specification, we use the failing test cases as counterexamples 
and the correct value  as its specification. 

Table \ref{tcas.tab} shows the result of running \ba on TCAS Test-suite. 
{\bf \ba ran 1440 times over all versions and 1367 of these runs 
pin-pointed the exact bug location, which is 95\% of the total runs}. 
The ``$TC\#$" in the table  is the number of failed test cases for each version. 
We ran \ba with each of these failing testcases as failing program executions and the 
golden output as the assertion to be satisfied. The column ``$Error\#$"  shows
the number of errors injected in to each version. Most versions have only 1 error 
but some have 2 and 3 errors. ``$Detect\#$" is the number of runs of \ba which detected
the correct (human-verified) bug location. ``$Size Reduc\%$" is the  percentage reduction in the 
code size given by the tool to locate the bug, the ratio of  bug locations returned by the tool 
to the total number of lines in the code. The ``$Run Time$'' shows the run time for
each run of \ba in seconds and they are  negligible.
The last column is the type of bug which is explained in Table \ref{tab.bugtype}.
 For example, the version v2 has one error injected and 
 has 69 failing testcases. We collected the bug locations reported during these 69 runs of 
the tool which gave 8 potential bug locations, which is 4.6\% of the 
total line number's in the program. The exact location of the fault is 
identified in all the 69 runs.

\begin{table}[htb]

\centering
\caption{Type of Error}
\label{tab.bugtype}
\begin{tabular}{||c|@{}l||}
\hline
\hline
Error Type & Explanation for the error \\
\hline
\hline
op    &  Wrong operator usage  \\
        &eg: $<=$ instead of $<$  \\
\hline
code  & Logical coding bug.  \\
\hline
assign &  Wrong assignment expression.  \\
\hline
addcode &   Error due to extra code fragments. \\ 
\hline
const  &  Wrong constant value supplied \\
           & eg: off-by-one error. \\
\hline
init  &    Wrong value initialization of a variable.  \\
\hline
index &   Use of wrong array index.  \\
\hline
branch &  Error in branching due to negation of \\ 
       &   branching condition  \\
\hline
\hline
\end{tabular}
\end{table}

Except for a few versions like
v12, v28 and v35, \ba detected the correct bug location for 
all the runs. For the remaining ones,
when we rank locations based on frequency of being reported as bugs,
exact bug locations had a count more than half of 
the total number of runs.  
The runs in which exact location was not reported did 
give clues about the real bug. For example, some testcases had  
wrong constant value assignment to an array element, 
for which the tool reported the fault at places where that array is accessed 
rather than the line at which the bad assignment occured.
By analyzing the error locations it is quite evident that the
error is due to a wrong value to that array location. 
On average the number of lines to check for potential bug is reduced to 8\% 
of the total code.  It should be noted that most of the single runs of the 
faulty version have captured the exact bug location. 


The Figure 2 gives an overview of a version of tcas (v2). with 
the bug at line~2, the original code is given in comment
in line~3. The declaration and initialization of 
variables,  functions and conditional statements that are not relevant 
to this bug are omitted in this example.  The bug is injected in  function 
$\textit{Inhibit\_Biased\_Climb}$ at line~2 by confusing the 
constant values. The original code is shown in comments at line~3.
 The program needs to satisfy the safety property  $\textit{alt\_sep\_test()}$ should
return $\textit{DOWNWARD\_RA}$ and is given as assertion at line~37.
There was 69 failing testcases for this version, we ran all these error traces
and the tool returned 8 potential bug locations which are shown underlined in Figure~2. 

There is no error reported in function $\textit{Non\_Crossing\_Climb(})$ because the 
call for that function at line~25 needs the function 
$\textit{Own\_Below\_Threat()}$ to be true, but that is false based on a comparison
on the input parameters which are made hard clauses. Now lets take a closer
look at the reported errors. 
\begin{itemize}
\item The line~34 is too weak for a fix because changing the return
value can make the assertion always true and that does not serve as a suitable
fix. 
\item In line~26 making the $\textit{need\_downward\_RA}$ variable
true can pick the right value for $\textit{alt\_sep}$. This decision is made 
by evaluation of the two functions in that statement. The 
 $\textit{Own\_Above\_Treat()}$ is true based on the input and it is clear that 
the correction needs to be done to the function call $\textit{Non\_Crossing\_Decend()}$.
\item The function $\textit{Non\_Crossing\_Decend()}$ has a call for the actual faulty
function at line~14. It also shows the repair could be 
done by changing the return value of this function at line~19, or where 
the wrong evaluation happens at (lines~15,16). 

\item The actual bug at line~2 is  reported as a
potential bug location in all the runs. 
It is interesting that all the other locations
were pointing to this line as the  base cause and helps the programmer to 
make a fix at the root cause of the problem. 

\end{itemize}

\begin{table*}[htb]
\centering
\label{tab.large}
\begin{tabular}{|c|c|c|c|c|c|c|c|c|c|c|c|c|}
\hline
 &Program &LOC\#&Proc\#&Reduc&\multicolumn{2}{|c|}{assign\#} &\multicolumn{2}{|c|}{var\#}& \multicolumn{2}{|c|}{clause\#} &Fault\#& time \\
 &    &  & & &Before&After&Before &After&Before& After & &\\
\hline
1&totinfo &  565 &  7   &   S  &  734 & 21  & 0.797m &400 & 1.822m & 1225 & 2& 0.19s  \\
2&print\_tokens& 726 & 18& C  & 65698  & 239   & 5.507m & 7439   & 53.483m  & 22634 & 13  & 25s \\ 
3&schedule& 564  & 21   &  DS  & 5914  & 391  & 5.173m& 0.053m & 15.379m&0.142m& 13 & 28s \\ 
4&schedule& 564  & 21   &  DS  & 41942  & 5412  & 78.982m& 4.517m & 239.385m&13.788m& 25 & 11h \\ 
5&totinfo &  565 &  7   &   CS  &  865 & 454  & 0.862m &0.734m & 4.156m &3.728m & 3 & 225s  \\
6&schedule2&  374 & 16   &  S  &  398  & 275  & 0.021m& 0.015m & 0.062m &0.048m & 9 & 20s  \\
\hline
\end{tabular}
\caption{Running Bug-Assist on larger benchmark programs from Siemens testsuite. } 
\end{table*}

\subsection{Larger Examples}
\label{exp.2}
To prove the scalability of our approach, and applicability in the presence
of complex pointers and loops, we choose a bunch of other 
testcases with function calls, recursion, dynamic memory allocation, loops, 
and complex programming constructs. 
In the TCAS testcases we did not apply 
any trace reduction method and used the entire boolean representation
of the program. When the program size and complexity increases, the 
error trace formula becomes huge. We do a preliminary investigation as proof of
concept on effectively reducing the error trace leveraging on the existing 
trace reduction techniques like program slicing (S), concolic simulation (C) and 
isolating failure-inducing input \cite{DDInput02} using delta debugging (D).

Table~3  shows the result of running \ba on 
4 other programs from the Siemens suite each with one injected fault.
``\textit{Program}" shows the name of the 
program from the Siemens testsuite. 
``\textit{LOC\#}" is the total lines of code  
in the program and ``\textit{Proc\#}", the number of procedure calls. 
The kind of reduction technique is specified in ''\textit{Reduc}" and  
``\textit{assign\#}" shows the size of the dynamic error trace as the number of 
assignment expressions before and after performing reduction technique. 
The ``\textit{var\#}" and ``\textit{clause\#}" is the number of 
boolean variables and clauses in the MAX-SAT representation of the 
error trace both before and after the reduction step mostly in millions(m).
The number of potential fault locations returned
by the tool is given under ``\textit{Fault\#}". 
The column ``\textit{Time}" shows the runtime in seconds(s) or hours(h).

We picked a faulty  version of the 
program and one test input that reveals the bug. 
 The golden output from the 
non-fault program with this same input is given as a post condition to this faulty program.
Trace reduction techniques are applied to the program execution with this input to 
generate a smaller trace formula and given as input to \ba. 
The tool reported the exact bug location in all programs except 
one (Program~2: \texttt{print\_token}).  
Trace reduction techniques significantly reduced the resulting
trace and the size of the MAX-SAT instance, as shown in ``Before" and ``After" 
sizes in Table~3. 
The cardinality of the potential fault location set for each of these  programs 
is very small.
In all cases, the run time of the tool is also smaller than the human effort required to isolate the fault
on the original trace.
This shows the applicability of the approach in complex real world programs.  

\begin{itemize}
\item 
The error inducing input to  Program~\texttt{totinfo} was  the rows and columns of a matrix. 
The bug was in the constant value of a conditional operator on checking the product of
rows and columns after a few other operations. A simple program slicing removed the 
assignments irrelevant to the assertion being checked and    
reduced the number of assignments to 21 with run time less than a second.

\item Program~\texttt{print\_token} contained a recursive function ``next\_token" and the input to the program
required the loops to be unrolled 8 times in the symbolic trace formula generation. 
This made the recursive function to have 64 instances in the symbolic trace and 
the number of assignments went up to 65K without concolic execution. 
Using concrete execution for the recursive function and variables 
brought down the number of assignment statements to 239.
It should be noted that the limitation in using a 
concrete execution would be to assume that the bug is not present in the functions and loops
which are concretized. 
However, this methodology fits well in  programs using functions from 
a reliable library or for functions which are already verified to be bug free. 
This program did not show the the bug at the exact location, which was a comparison on 
a variable which got the value from the concrete execution. 
This was because the constant propagation used by the symbolic trace generator abstracted away the variable
since its values was a constant. 
Instead, the error was shown in the assignment of the variable to the constant.  

\item The priority scheduler program~3 and 4, contained a large error inducing input which 
called a bunch of procedures before deviating from the golden output of the original program. 
The trace size was significantly reduced after isolating the error inducing input using delta debugging,
but was still quite big (about 400 and 5400 assignment operations respectively).
In program~3, the off-by-one error on flushing the number of processes 
was detected by the presence of a single process creation (leading to a trace of about 400 assignments). 
But program~4 required a much larger 
input and more procedures to expose the failure, resulting in a longer trace. 
It took \ba almost 11 hours to find the exact location (excluding the time taken for input 
minimization using delta debugging). 
Each execution of MAX-SAT took around 30 minutes to identify one potential 
fault location.  
\end{itemize}

\subsection{Fixing Off-By-One Errors}
\begin{code}
\begin{Verbatim}[commandchars=\\\{\},
                codes={\catcode`$=3\catcode`^=7}]

1   #define SIZE 15
2   void MyFunCopy (char *s) 
3   \{
4       char buf[SIZE];
5       memset(buf, 0, SIZE);
6       \bug{strncat(buf, s, SIZE);} 
7      /*Last argument should be: SIZE-1 */
8       return;
9   \}

/*Standard C implementation of strncat*/
10   char *strncat(char *dest, const char *src, 
                                    size\_t n)
11   \{
12         char *ret = dest;
13          while (*dest)
14             dest++;
15          while (n--)
16              if (!(*dest++ = *src++))
17                  return ret;
18          *dest = 0; /*Problem cause*/
19          return ret;
20    \}
\end{Verbatim}
\caption{The strncpy program with Off-By-One error.}
\label{code.offbyone}
\end{code}
We demonstrate the repair capability of \ba by synthesizing fixes for 
Off-By-One error which are common programming error for users of  C 
library routines because of their
inconsistency with respect to whether one needs to subtract one byte or 
use the correct size. One common Off-By-One error in C library  which 
results in security related theart is caused by the misuse of  
$strncat$ routine \cite{Todd99}. 
A common misconception with strncat is that 
the guaranteed null termination will not write beyond the maximum length. 
In reality it will write a terminating null character one byte beyond the 
maximum length specified.

The Program \ref{code.offbyone} shows an instance of the bug in the function 
$MyFunCopy$, which takes a string $s$ and uses the $strncat$ routine to copy 
the contents to a string $buf$ of length $SIZE$. The lines~10--20 shows a standard {C} implementation of strncat.
Note that after copying the $n$ characters at line~17 
it writes to the $n+1^{th}$ location of the $dest$ string at 
line~18. This require that the function $MyFunCopy()$ should
be using $SIZE-1$ as the last argument to function $strncat$.  

We ran \ba on this function turning on the check for accesses within array bounds.
It located the line~6 as a potential bug
location in the code. We have taken the assumption that the library functions
cannot be modified and in the {pMAX-SAT} problem formulation we make 
constraints arising out of library functions hard clauses. 
This location is already marked during preprocessing as a statement with a
constant; the \ba now tries to fix it by changing the value to $SIZE-1$ and 
$SIZE+1$ as explained in the Algorithm \ref{fig.offbyone}. 
This requires turning off constant propagation while converting the 
program in to boolean formula and collecting the literals in the {CNF} 
corresponding to each constant. Then we create two {SAT} instances with these 
new constant values and give it to MiniSAT solver to check property 
violations. In this example it came up with a success on the value $SIZE-1$ 
and is provided as a fix for the fault. 
 
\subsection{Finding Faulty Loop Iteration}

\begin{code}
\begin{Verbatim}[commandchars=\\\{\},
                codes={\catcode`$=3\catcode`^=7}]
1    int squareroot()
2    \{
3        int val = 50;
4        int i =1;
5        int v =0;
6        int res =0;
7        while(v < val)
8        \{
9            \bug{v = v + 2*i +1;}
10           \bug{i = i+1;}
11       \}
12        \bug{res = i;}
13        \codecomment{/* res = i - 1; */}
14       assert( (res*res <= val) && 
                    ((res+1)*(res+1) > val);
15         return res;
    \}
\end{Verbatim}
\caption{The nearest integer square root function with bug at line~12}
\label{code.squareroot}
\end{code}

The program \ref{code.squareroot} contains a function to find the nearest 
integer square root of a value.   The post condition specified 
as assertion requires that the $res$  should be the closest square root 
for $val$. The  bug locations reported by \ba are underlined. The 
correct code is given as comment in line~13.  Even though the actual 
bug is not in the loop body it requires a through analysis of the loop 
to conclude the right fix  at line~12. We gave the unwinding limit
$50$ to {CBMC} and the \ba reports the $8^{th}$ iteration of the  loop 
as the first occurrence of line~10 fault.

\section{Scalability and Limitations}
The fault localization depends on the underlying 
boolean transform of the program to clauses. 
Therefore the code omission faults cannot be detected due to the 
non existance of those clauses, instead it tries to fix expressions with in the 
current program to validate the asserted property.
In most of the cases a single error trace was sufficient to locate 
the exact error location and that shows the speed up of this method 
compared to the existing fault localization approaches. Each of the 
potential error locations are the unsatisfied clauses  during each iteration 
of the MAX-SAT solver. Using an incremental SAT approach for each of 
these iterations can considerably bring down the running time of the tool.
Moreover, there is a growing interest in extracting
the unsatisfiable cores \cite{msc} which can further aid this approach.   

As shown in the experimental results, without applying any 
trace reduction technique this method may blow up the state space 
and may not be suitable for programs with complex calls or 
enormous lines of code.
However, the tool would be handy in an Integrated Development 
Environment(IDE) where the programmer is interested  in debugging the function under
development and can  abstract away the rest of the program as 
input output relationship.
This methodology can provide online hints for the programmer assisting in  
code development phase and is the motivation in developing the Eclipse plugin
for the tool.  Any error trace reduction method can also be applied
orthogonally to this approach to bring down the trace.

\section{Conclusions}
\label{sec-conclusions}

Program analysis based on Boolean satisfiability has been
extremely successful in detecting subtle errors in large
software programs \cite{cbmc,XieAiken,EXE}.
We show that techniques based on Boolean MAX-SAT can be similarly
effective in {\em localizing} program errors (as well as in identifying
potential fixes). 

Our technique can leverage engineering advances in modern SAT and
MAX-SAT solvers, and as our experiments demonstrate, provide a precise
and scalable solution to the error localization problem.
While we have described error localization at the line-number (or program statement)
level, our reduction to {pMAX-SAT} is general, and can be used at different
levels of granularity. 
For example, to localize bugs at the module level, we can group 
clauses coming from the same module in the {pMAX-SAT} instance. 

To improve the usability of our tool, we have built an Eclipse plugin 
to help the programmer to find 
bug locations during the development process. It marks the potential 
bugs in the code under development  and assist in analyzing the
right fix. The tool also marks the repair capabilities at a line and 
the user can also ask for automated repair like Off-By-One fix as discussed 
in this paper.

In future we would like to explore the various automated bug fixing 
capabilities by analyzing the bug locations. This requires predicting the type 
of error which has a maximum probability in a particular expression. 
It would be interesting to mine the software repositories for bug patterns and
building a model for expression specific  error types based on the
repository history and use it for guiding \ba for an appropriate repair
strategy.
Another direction is to provide constructive suggestions to the programmer 
in  fixing a bug.  For example, Suppose the \ba comes up with an error statement
 which has a constant; showing the lower and upper bound of the values 
for that constant which holds the given properties 
help the programmer to provide a robust fix.

\bibliographystyle{plain}
\bibliography{sw}

\end{document}